\documentclass{webofc}
\usepackage[varg]{txfonts}   
\usepackage{slashed}

\begin{document}
\title{Polarization in heavy ion collisions: a theoretical review}
%
%

\author{\firstname{Matteo} \lastname{Buzzegoli}\inst{1}\fnsep\thanks{\email{mbuzz@iastate.edu}\\ Presented at the 20th International Conference on Strangeness in Quark Matter (SQM 2022)}
}

\institute{Department of Physics and Astronomy, Iowa State University, Ames, Iowa 50011, USA}

\abstract{%
  In these proceedings I discuss the recent progress in the theory of spin polarization in relativistic fluids.
  To date, a number of studies have begun to examine the impact of the shear tensor on the local spin polarization
  and whether this contribution can restore agreement between the measurements and the predictions obtained from
  a polarization induced by the gradients of the plasma. I present the derivation of the spin polarization vector of a fermion at local thermal
  equilibrium and I discuss the role of pseudo-gauge transformations and of dissipative effects. I list what
 we can learn from the polarization measured at lower energies. Finally, I discuss possible applications of spin
  polarization measurements in relativistic heavy ion collisions.
}
\maketitle
\section{Introduction}
\label{intro}
Particles in a rotating medium tends to align their spin in the direction of the total angular momentum
due to the spin-rotation coupling. It was then expected that fermions produced in relativistic heavy-ion
collisions will have a global spin polarization proportional to the vorticity of the fluid \cite{Liang:2004ph,Becattini:2007nd,Becattini:2007sr}.
Based on this idea, one obtains that the spin polarization vector of a fermion is given by a Cooper-Frye like formula~\cite{Becattini:2013fla}:
\begin{equation}
\label{eq:SpinPolVorticity}
S^\mu(p) = - \frac{1}{8m}\epsilon^{\mu\rho\sigma\tau} p_\tau 
  \frac{\int_{\Sigma} {\rm d} \Sigma \cdot p \; n_F (1 -n_F) \varpi_{\rho\sigma} }{\int_{\Sigma} {\rm d} \Sigma \cdot p \; n_F},
\end{equation}
where
\begin{equation}
\varpi_{\mu\nu}=-\frac{1}{2}\left(\partial_\mu\beta_\nu - \partial_\nu\beta_\mu \right)
\end{equation}
is the thermal vorticity tensor and $n_F=\left({\rm e}^{\beta\cdot p-\mu/T}+1 \right)^{-1}$  is the Fermi distribution function with
$\beta^\mu=u^\mu /T$ the four-temperature vector and $\mu$ a chemical potential. The predictions for global polarization obtained
from this formula were confirmed in 2017 with $\Lambda$ hyperons produced in Au-Au collisions at $\sqrt{s_{NN}}= 200$ Gev \cite{star,Adam:2018ivw},
see for instance \cite{Becattini:2020ngo} for a review.

However, models based on the vorticity induced spin polarization (\ref{eq:SpinPolVorticity}) are not able to
reproduce later measurements of local $\Lambda$ spin polarization, that is as a function of the particle momentum \cite{Adam:2019srw,Niida:2018hfw,ALICE:2021pzu}.
This is known as local polarization sign puzzle because the models predicted the opposite sign for
the longitudinal (along the beam axis) component of the spin polarization vector. Once it was established that this discrepancy
can not be explained with feed-down corrections \cite{Xia:2019fjf,Becattini:2019ntv}, the sign puzzle
was a strong motivation for theoretical investigation as something relevant was clearly missing in the
theoretical description. Further investigations, that I review in next section, indeed revealed that spin
polarization can also be induced by the shear flow of the fluid and, for heavy-ion collisions, this turned
out to be a relevant contribution.

\section{Spin polarization at local thermal equilibrium}
In this section I review the derivation of the spin polarization vector
of a fermion in a fluid that reached the local thermal equilibrium.
The spin polarization vector is obtained from the particle branch
of the Wigner function using (see the review \cite{Becattini:2020sww} for details)
\begin{equation}
\label{eq:SpinPolFormula}
S^\mu_\varpi(p)= \frac{1}{2}\frac{\int_{\Sigma_{FO}} {\rm d}\Sigma\cdot p\; {\rm tr}_4\left[\gamma^\mu \gamma^5 W_+(x,p) \right] }
 {\int_{\Sigma_{FO}} {\rm d}\Sigma\cdot p\; {\rm tr}_4 \left[W_+(x,p)\right]},
\end{equation}
where, for heavy-ion collision applications, the hypersurface $\Sigma_{FO}$ is
the decoupling hypersurface, where the system can be described with the quasi-free
hadronic effective fields. Hence, $\Lambda$ particles being quasi-free,
we can use the Wigner function of a free Dirac field:
\begin{equation}
\label{eq:WignerFunc}
W(x,k)_{AB} = \frac{1}{(2\pi)^4} \! \int \!{\rm d}^4 y\, {\rm e}^{-{\rm i} k \cdot y}
	\langle : \bar{\Psi}_B (x +y/2) \Psi_A (x-y/2) : \rangle,
\end{equation}
where the brackets denotes the thermal average with the statistical operator $\hat{\rho}$:
$\langle \hat{X} \rangle = {\rm tr}\left(\hat{\rho}\hat{X}\right)$.

For a relativistic system whose underlying microscopic theory is a quantum field theory, the
statistical operator describing the thermal states can be obtained using the Zubarev method \cite{Becattini:2019dxo}.
By maximazing the entropy of the system at the moment of thermalization and imposing constrains
on the momenta and energy densities, one obtains the covariant statistical operator at the time
of decoupling
\begin{equation}
\label{eq:StatOperZubarev}
 \widehat{\rho} = \frac{1}{Z} \exp\Bigg[
-\int_{\Sigma_{FO}} {\rm d}\Sigma_\mu\left(\widehat{T}_B^{\mu\nu}\beta_\nu
 -\zeta\,\widehat{j}^\mu\right)
+\int_\Theta {\rm d}\Theta\left(\widehat{T}_B^{\mu\nu}\nabla_\mu\beta_\nu
 -\widehat{j}^\mu \nabla_\mu \zeta \right)\Bigg],
\end{equation}
where $\widehat{T}_B^{\mu\nu}$ is the Belinfante stress-energy tensor (SET) operator, $\widehat{j}$ is
a conserved current, and $\zeta=\mu/T$.
One can show that the second term in the exponent gives rise to dissipative effects, while the
first term gives the non-dissipative local thermal equilibrium. I postpone the discussion of the
dissipative part in a later section and here I focus on the local equilibrium statistical operator:
\begin{equation}
\label{eq:LEStatOper}
\widehat{\rho}\simeq \widehat{\rho}_{{\rm LE}}
	= \frac{1}{Z} \exp\left[ -\int_{\Sigma_{FO}}\!\! {\rm d}\Sigma_\mu
	\left( \widehat{T}_B^{\mu\nu}\beta_\nu -\zeta\,\widehat{j}^\mu\right)\right].
\end{equation}
The derivation of the spin polarization at local thermal equilibrium from Eq. (\ref{eq:SpinPolFormula})
reduces to the evaluation of the Wigner function at local equilibrium
\begin{equation}
\label{eq:WignerLE}
W(x,k)_{LE}= \frac{1}{Z} {\rm tr} \left( \exp\left[
-\int_{\Sigma_{FO}}\!\! {\rm d}\Sigma_\mu(y)\left(\widehat{T}_B^{\mu\nu}(y)\beta_\nu(y)
 -\zeta(y)\,\widehat{j}^\mu(y)\right)\right] \widehat{W}(x,k) \right).
\end{equation}
For a fluid in the hydrodynamic regime, a good approximation of (\ref{eq:WignerLE}) is obtained
expanding the thermodynamic fields as a series of gradients around the point $x$ and using
the linear response theory to obtain the Wigner function mean value. Keeping the first order in derivatives of
the four-temperature and of the chemical potential
\begin{equation}
\label{eq:TaylorBeta}
\beta_\nu(y) \simeq \beta_\nu(x) + \partial_\lambda \beta_\nu(x) (y-x)^\lambda,\quad
\zeta(y) \simeq \zeta_\nu(x) + \partial_\lambda \zeta(x) (y-x)^\lambda\, ,
\end{equation}
one finds
\begin{equation}
\label{eq:ApproxHydro}
\begin{split}
\widehat{\rho}_{LE} \simeq \frac{1}{Z} \exp& \left[ - \beta_\nu(x) \widehat{P}^\nu +\zeta(x)\widehat{Q}
	+ \frac{1}{2} \varpi_{\mu\nu}(x) \widehat{J}^{\mu\nu}_x  -\frac{1}{2} \xi_{\mu\nu}(x) \widehat{Q}^{\mu\nu}_x\right.\\
	&\left.+\partial_\lambda \zeta(x)\!\!\! \int\!\!\! {\rm d} \Sigma_\mu (y-x)^\lambda \widehat{j}^{\mu}(y)+\cdots\right],
\end{split}
\end{equation}
where
\begin{equation}
\widehat{J}^{\mu\nu}_x =\! \int \!{\rm d} \Sigma_\lambda \left[ (y-x)^\mu \widehat{T}_B^{\lambda\nu}(y) -
  (y-x)^\nu \widehat{T}_B^{\lambda\mu}(y)\right],
\end{equation}
is the conserved angular momentum operator (the generator of boosts and rotations) and
\begin{equation}
\widehat{Q}^{\mu\nu}_x =\! \int \!{\rm d} \Sigma_\lambda \left[ (y-x)^\mu \widehat{T}_B^{\lambda\nu}(y) + 
  (y-x)^\nu \widehat{T}_B^{\lambda\mu}(y)\right],
\end{equation}
is a non-conserved symmetric quadrupole like operator. Notice that $\widehat{J}^{\mu\nu}$ couples with the thermal vorticity
$\varpi$, while $\widehat{Q}^{\mu\nu}$ couples with the thermal shear tensor defined as
\begin{equation}
\xi_{\mu\nu}=\frac{1}{2}\left(\partial_\mu\beta_\nu + \partial_\nu\beta_\mu \right).
\end{equation}
The spin polarization vector obtained from (\ref{eq:SpinPolFormula}) using the linear response theory
on the operators appearing in (\ref{eq:ApproxHydro}) is
\begin{equation}
\label{eq:SpinPolFirstOrder}
S^\mu(p) = - \epsilon^{\mu\rho\sigma\tau} p_\tau 
  \frac{\int_{\Sigma} {\rm d} \Sigma \cdot p \; n_F (1 -n_F) 
  \left[ \varpi_{\rho\sigma} + 2\, \hat t_\rho \frac{p^\lambda}{\varepsilon} \xi_{\lambda\sigma}
	-\frac{\hat{t}_\rho\partial_{\sigma}\zeta}{2\varepsilon}\right]}
  { 8m \int_{\Sigma} {\rm d} \Sigma \cdot p \; n_F},
\end{equation}
where $\hat{t}$ is the time direction in the laboratory frame. The first term is the vorticity induced polarization
(\ref{eq:SpinPolVorticity}), the second is the recently found shear induced polarization \cite{Becattini:2021suc,Becattini:2021iol,Liu:2021uhn,Liu:2021nyg,Yi:2021ryh},
and the last one is the contribution from the gradient of chemical potential obtained in \cite{Liu:2020dxg,Fu:2022myl}, sometimes called spin Hall effect term.

\subsection{Thermal shear and solution of the sign puzzle}
The effect of thermal shear in spin polarization was studied in ref. \cite{Becattini:2021iol,Yi:2021ryh,Fu:2021pok,Ryu:2021lnx,Florkowski:2021xvy,Sun:2021nsg,Alzhrani:2022dpi,Wu:2022mkr}.
These analyses revealed the following facts. The thermal shear tensor at the freeze-out hypersurface can be larger than the thermal vorticity. Because of its momentum
dependence (see Eq. (\ref{eq:SpinPolFirstOrder})), the thermal shear does not contribute to global polarization but it was found to give a large
contribution to local spin polarization. Predictions including the thermal shear are closer to the experimental data but they are still lacking
a good quantitative agreement.

In order to reproduce the experimental data, no new effects are needed. It is sufficient to improve the approximation used to derive
the spin polarization spectrum (\ref{eq:SpinPolFirstOrder}). Since the available experimental data for local spin polarization
was taken at 200 GeV and at 5 TeV, at such high energies it is expected that the hadronization happens at the same temperature.
In that case the statistical operator is given by the Isothermal Local Equilibrium operator:
\begin{equation}
\label{eq:ILEStatOper}
\widehat{\rho}_{\rm ILE} = \frac{1}{Z} \exp\left[
-\int_{\Sigma_{FO}}\!\! {\rm d}\Sigma_\mu \widehat{T}_B^{\mu\nu}\left( \frac{u_\nu}{T_{\rm FO}}\right) \right]
=\frac{1}{Z} \exp\left[-\frac{1}{T_{\rm FO}}\int_{\Sigma_{FO}}\!\! {\rm d}\Sigma_\mu \widehat{T}_B^{\mu\nu}u_\nu \right],
\end{equation}
because the temperature $T_{\rm FO}$ is constant along the decoupling hypersurface. In this setting, expanding the temperature
in a Taylor series together with $\beta$ as in Eq. (\ref{eq:TaylorBeta}) will only introduce unnecessary mistakes.
The appropriate spin polarization spectrum at high energies involves only the fluid velocity gradients \cite{Becattini:2021iol}:
\begin{equation}
\label{eq:ILEPol}
S_{\rm ILE}^\mu(p) = - \epsilon^{\mu\rho\sigma\tau} p_\tau 
  \frac{\int_{\Sigma} {\rm d} \Sigma \cdot p \; n_F (1 -n_F) 
  \left[ \omega_{\rho\sigma} + 2\, \hat t_\rho \frac{p^\lambda}{\varepsilon} \Xi_{\lambda\sigma} \right]}
  { 8m T_{\rm FO} \int_{\Sigma} {\rm d} \Sigma \cdot p \; n_F},
\end{equation}
where
\begin{equation}
\label{eq:Decompositions}
\omega_{\rho\sigma} = \frac{1}{2} \left(\partial_\sigma u_\rho - \partial_\rho u_\sigma \right),\quad
\Xi_{\rho\sigma} =  \frac{1}{2} \left(\partial_\sigma u_\rho + \partial_\rho u_\sigma \right)
\end{equation}
denote the kinematic vorticity and shear tensors. The predictions of (\ref{eq:ILEPol}) are in quantitative
agreement with the data taken at 200 GeV \cite{Becattini:2021iol}. Preliminary results presented at this
conference \cite{Palermo:sqm22} showed an improvement of the agreement when including the feed-down effects
and that the predictions of (\ref{eq:ILEPol}) are also in quantitative agreement with the local spin polarization
measured by ALICE \cite{ALICE:2021pzu}. Future analyses of high energy spin polarization should use the
isothermal settings.

What also emerged from these analyses is that the local spin polarization is sensitive to the properties of the
plasma, which ultimately determines the gradients of the thermo-hydrodynamic fields in (\ref{eq:SpinPolFirstOrder}).
This feature allows to use spin polarization as a probe of the quark gluon plasma properties.

\subsection{Low energies}
With new data of spin polarization at very low energy \cite{STAR:2021beb}, we can further test the models and
see whether they can reproduce the maximum of spin polarization \cite{Ivanov:2019ern,Ivanov:2022ble,Deng:2020ygd,Deng:2021miw,Guo:2021udq}
expected to occurs around $\sqrt{S_{NN}}=7$ GeV. 
This energy range is also important for polarization in the FAIR and NICA facilities.
At lower energies we also observe a larger deviation between the polarization of $\Lambda$ and
$\bar{\Lambda}$, which allows to probe the gradients of baryonic chemical potential \cite{Liu:2020dxg,Fu:2022myl} and
the effect of the magnetic field \cite{Guo:2019joy}. Physics of rotating matter at these energies was also discussed in
this conference \cite{Liao:sqm22}.

\subsection{Spin tensor and pseudo-gauge dependence}
For the first time, the importance of the spin degrees of freedom in the determination of the spin polarization vector,
opens the possibility to have a direct experimental test of a local description of spin, i.e. a spin tensor.
Indeed one can include a spin tensor in hydrodynamic equations \cite{Florkowski:2018fap,Hattori:2019lfp,Hongo:2021ona,Hongo:2022izs,Bhadury:2020cop,Bhadury:2021oat,She:2021lhe}
and let it evolve with the rest. In the same way that the temperature is a thermodynamic potential for energy, the
thermodynamic quantity associated with spin is called spin potential: $\Omega_{\mu\nu}$, and it might be present
in spin polarization.

The definition of spin tensor is not unique in special relativity but is given up to a transformation
called pseudo-gauge transformation. For instance, from the Belinfante pseudo-gauge with SET $\widehat{T}^{\mu\nu}_B$ and vanishing spin
tensor, a generic pseudo-gauge $\Phi$ is obtained with:
\begin{equation}
\begin{split}
	\widehat{T}^{\mu\nu}_\Phi=& \widehat{T}^{\mu\nu}_B+\frac{1}{2}\nabla_\lambda\left(\widehat{\Phi}^{\lambda,\mu\nu}
	-\widehat{\Phi}^{\mu,\lambda\nu} -\widehat{\Phi}^{\nu,\lambda\mu}\right),\\
\widehat{\mathcal S}^{\lambda,\mu\nu}_\Phi=& -\widehat{\Phi}^{\lambda,\mu\nu}+ \nabla_\rho\widehat{Z}^{\mu\nu,\lambda\rho},
\end{split}
\end{equation}
where $\widehat{\Phi}^{\lambda,\mu\nu}=-\widehat{\Phi}^{\lambda,\nu\mu}$ and
$\widehat{Z}^{\mu\nu,\lambda\rho}=-\widehat{Z}^{\nu\mu,\lambda\rho}=-\widehat{Z}^{\mu\nu,\rho\lambda}$.
This raises the question whether a pseudo-gauge transformation affects the results for spin polarization.
It was indeed realized that the local equilibrium form of statistical operator is different for different
pseudo-gauges \cite{Becattini:2018duy}:
\begin{equation}
\label{eq:RhoPhi}
\widehat{\rho}_{\rm LTE}^{\,\Phi}=\frac{1}{\mathcal{Z}}\exp\left\{\!-\!\int\!\!{\rm d}\Sigma_\mu\!\!\left[ \widehat{T}^{\mu\nu}_{B}\beta_\nu\!
	-\!\frac{1}{2}\left(\varpi_{\lambda\nu}-\Omega_{\lambda\nu}\right)\widehat{\Phi}^{\mu,\lambda\nu}
	- \xi_{\lambda\nu} \widehat{\Phi}^{\lambda,\mu\nu}-\frac{1}{2}\Omega_{\lambda\nu}\nabla_\rho\widehat{Z}^{\lambda\nu,\mu\rho}-\widehat{j}^\mu\zeta\right]\right\}.
\end{equation}
Starting from this, the different predictions of the spin polarization spectrum for several pseudo-gauge
were obtained \cite{Buzzegoli:2021wlg}. This difference is not only related to the contribution of the spin potential but also
of the thermal shear. This feature is not unique of the statistical mechanics approach
and the role of pseudo-gauge was studied in different approaches \cite{Speranza:2020ilk,Fukushima:2020ucl,Das:2021aar,Li:2020eon,Weickgenannt:2022jes}.
This is an unique opportunity to study the physical meaning of pseudo-gauge transformations and possibly to shed
light on the role of spin potential in gravitational phenomena.

\section{Dissipative effects and relativistic kinetic theory with spin}
The current data and the inclusion of the thermal shear support the local equilibrium picture with a small
contribution of dissipative effects. However, one should check how much the dissipative phenomena affect the spin polarization.
These corrections in the spin polarization formula have been found for the massless \cite{Shi:2020htn}
and massive \cite{Weickgenannt:2022zxs} Dirac field, but so far there are no analyses on the impact of these
effects on the spin polarization of $\Lambda$ particles in heavy-ion collisions. If the dissipative effects will be
found to be relevant, this would imply a separation of scales between the interaction time and the time in which the spin
degrees of freedom equilibrates \cite{Hongo:2021ona,Hongo:2022izs,Hidaka:2017auj}. In this scenario the inclusion
of a spin tensor in hydrodynamics mentioned above will be crucial.

An estimate of the magnitude of the dissipative effect might not be easy, as it was realized that, in addition to usual transport
coefficients, in a hydrodynamic with spin additional transport coefficients related to spin emerge \cite{Weickgenannt:2022zxs,Hongo:2021ona,Hongo:2022izs,Hidaka:2017auj}.
The values of these new transport coefficients for the QGP can be fixed by fitting the data of spin polarization,
but in this way one inevitably loose the predictive power of the current theory.

A powerful approach to study dissipative effects and the role of spin in hydrodynamics is the relativistic kinetic theory.
The Wigner-function formalism in particular is convenient to preserve all the covariant properties of the
system and to find results as an expansion in $\hbar$. Motivated by spin polarization, this approach recently
experienced a rich and fast development \cite{Weickgenannt:2019dks,Gao:2019znl,Hattori:2019ahi,Liu:2020flb,Weickgenannt:2020aaf,Weickgenannt:2021cuo}.

In this regard, I would like to draw attention to an often overlook assumption needed to solve
the Wigner equation in kinetic theory with collisions:
\begin{equation}
\label{eq:WignerEq}
\left[ \gamma\cdot\left(p + {\rm i}\frac{\hbar}{2}\partial\right) -m \right]
	W_{\alpha\beta} = \hbar\, \mathcal{C}_{\alpha\beta}.
\end{equation}
Wigner equation itself is not enough to determine all the properties of the system.
For instance, changing $\widehat{\Phi}$ in (\ref{eq:RhoPhi}) results in different Wigner functions all solving Eq. (\ref{eq:WignerEq}).
Indeed the determination of the non equilibrium, dissipative Wigner function from (\ref{eq:WignerEq}) requires the input of the equilibrium Wigner function.
For a system in absence of vorticity, the equilibrium form is well known. But the general form of global equilibrium Wigner function in the presence
of vorticity was obtained only recently \cite{Becattini:2020qol,Palermo:2021hlf}. Usually the ansatz for the equilibrium form used in kinetic theory
agrees with the correct form only at first order in thermal vorticity. For the current applications is enough, but one have to keep it in mind when
deriving higher order results.

\section{Applications of spin polarization in heavy ion collisions}
Spin is a new element in heavy-ion collisions and as a commonly accepted
theoretical framework is being built, several ideas on how to use it as a
diagnostic tool for the QGP have been proposed. What makes spin polarization
a good probe to the QGP properties is its dependence on the gradients of
the thermo-hydrodynamics fields (fluid velocity, temperature and baryonic chemical potential).

For instance, reproducing the vorticity structure of the plasma requires
knowledge of the properties and evolution of the QGP. It was then proposed
to use spin correlations of two $\Lambda$ hyperons to probe the vorticity structure
of the fluid \cite{Pang:2016igs,Xia:2018tes,Ryu:2021lnx}. In this way one acquires
more constraints on the properties of the QGP.

Correlations of helicity, that is the projection of spin along the momentum of the particle,
of two hyperons in the same event can also reveal the presence of local parity violation
in hot QCD matter \cite{Du:2008zzb,Becattini:2020xbh,Gao:2021rom}. Local parity violations
should create an imbalance of right-handed and left-handed quarks in the plasma, described
with an axial chemical potential $\mu_A$. Currently the chiral magnetic effect is used
to reveal the presence of $\mu_A$ \cite{Kharzeev:2004ey}. But the axial imbalance
should also add a contribution to spin polarization (\ref{eq:SpinPolFirstOrder}) as \cite{Becattini:2020xbh}
\begin{equation}
S^\mu(p) = S^\mu_{\rm hydro}(p) + \frac{g_h}{2} \frac{\int_\Sigma {\rm d}\Sigma\cdot p \; (\mu_A/T) n_{F}\left(1-n_{F}\right)}{\int_\Sigma {\rm d}\Sigma \cdot p \; n_{F}} \frac{\varepsilon p^\mu- m^2 \hat t^\mu}{m \varepsilon},
\end{equation}
resulting into a peculiar (parity breaking) contribution to helicity correlations.

Spin polarization can be used to study anything that produce large enough gradients.
This is the case of a jet crossing the plasma. The vorticity field generated by it contains
information about the energy lost by the jet and can be studied with spin polarization
measurements in reference to the jet axis \cite{Serenone:2021zef}.

The results derived so far for spin polarization do not include interactions.
Vorticity induced polarization is a consequence of spin-rotation coupling $g_\Omega \vec{S}\cdot \vec{\Omega}$,
and Einstein Equivalence Principle (EEP) prevents spin-rotation coupling to receive
radiative corrections: $g_\Omega\equiv 1$ \cite{kobzarev1962gravitational}. However, the
breaking of Lorentz covariance at finite temperature also breaks the EEP and this allows
interactions to affect the spin-rotation coupling \cite{Buzzegoli:2021jeh}. For strong interactions
the leading order correction is expected to be
\begin{equation}
g_\Omega=1-\frac{N_c^2-1}{2}\frac{1}{6}\frac{g^2 T^2}{m^2},\quad T\ll m 
\end{equation}
which, in a typical event, results in a reduction of about 40\% in the spin polarization of the strange quark.
Whether we can detected this consequence of the breaking of EEP in the energy dependence
of $\Lambda$ polarization is still under investigation.
 

\section{Conclusions}
Spin polarization is a new topic in heavy ion collisions that just started to show
its potential. It can be used to answer many questions left unanswered in relativistic
hydrodynamics and to probe the properties of hot QCD matter. When all the relevant
gradients are included and more accurate approximations adopted, the description of
the QGP as a fluid close to local equilibrium with small viscous corrections is
receiving confirmation as a solid model in heavy ion collisions. Right now, theory
and experiments in this field are rapidly expanding and spin will be used as a tool
to investigate fundamental physics in the QGP and beyond.

\vskip0.3cm
\textbf{Acknowledgments.} M.B. is supported by the US Department of Energy under Grant No. DE-FG02-87ER40371.

\end{document}